# Algorithmic Governance for Explainability: A Comparative Overview of Progress and Trends

**Yulu Pi**
University of Warwick
yulu.pi@warwick.ac.uk

## Abstract

The explainability of AI has transformed from a purely technical issue to a complex issue closely related to algorithmic governance and algorithmic security. The lack of explainable AI (XAI) brings adverse effects that can cross all economic classes and national borders. Despite efforts in governance, technical, and policy exchange have been made in XAI by multiple stakeholders, including the public sector, enterprises, and international organizations, respectively. XAI is still in its infancy. Future applications and corresponding regulatory instruments are still dependent on the collaborative engagement of all parties.

## 1 Introduction

In 2020, due to the COVID-19 pandemic, the UK government canceled the A-level exam and instead used an AI model to predict student performance. The model systematically estimated that students from minority and low-income communities had lower grades than their historical grades, making these students less likely to get into prestigious universities. Despite the controversy, the UK government was unable to explain why the algorithm produced such systematic discrimination and ultimately had to eventually switch to the manual estimation of student grades by teachers.

This incident highlights the social problems that arise from the use of algorithms that lack transparency and explainability. The lack of clear explanations for the algorithm's outcomes can lead to unjust and discriminatory practices, which can have far-reaching consequences. As a result, policymakers and stakeholders have been working to address these issues by implementing policy measures and developing technical solutions to ensure the explainability of AI algorithms.

The paper is a work in progress to provide a comparative review of the growing policy efforts to promote the explainability of AI algorithms. Through this analysis, the paper aims to contribute to the ongoing discussions on the governance of AI and the need for greater transparency and accountability in algorithmic decision-making.

## 2 Explainability challenge of AI

Human intelligence and AI are complementary. AI can extend human cognition to deal with complexity, while humans can still provide more holistic and intuitive thinking to deal with uncertainty. The complementarity of humans and AI can combine humans intuitive and holistic thinking with AI's super performance in collecting and analyzing information together. This human-AI collaboration is, however, significantly hampered by the "black box" nature of AI, which is the inability of humans to comprehend the algorithm's decision-making process. A common workflow for human-AI symbiosis in decision-making is for the AI to jump in first to analyze the problem and provide its judgment to wait for the human's final decision. For example, by sifting through and analyzing gigabytes of user-generated data, AI bots detect inappropriate or controversial websites or social media material, while human professionals "behind the AI curtain" use their judgment and make the final decision to remove a social media post or video(Jarrahi,2018). Humans' involvement is key to better manage the use of AI and prevent the jeopardization it causes. However, many AI algorithms, especially machine learning algorithms represented by deep neural networks with a vast number of parameters, are 'black box' in nature. The lack of access to the inner workings of algorithms makes it impossible for humans to trust AI, which impedes such cooperation as well as scrutiny of AI.

AI that lacks explainability impedes human-machine collaboration and "makes it difficult to apply AI on a large scale in fields such as defense, finance, healthcare, law, cyber security, etc., where stakes are high, or in applications that require legal compliance." Even more worrying is that AI that lacks explainability may bring issues related to algorithmic discrimination, algorithmic safety and algorithmic liability. Machine learning models have the supreme ability to learn from vast amounts of data. Through data rather than specific programming, machine learning models can autonomously construct rules of operation and achieve autonomous learning as well as self-iteration, allowing a single model to be applied to multiple scenarios and greatly freeing the productivity of AI engineers. However, the exact rules that machine learning models construct from their input data to their output results are not fully understood even by expert users. Such opacity prevents humans from understanding how AI makes certain decisions, making it impossible for humans to detect and correct errors, biases, discrimination, and other problems generated by AI. There is growing evidence that the widespread use of AI is often accompanied by adverse consequences such as bias against specific groups and anomalous errors that are difficult to detect. COMPAS, a crime risk assessment AI once widely used in the United States, has been shown to systematically discriminate against African-Americans. Applying such a biased algorithmic system to critical decision-making areas such as finance, healthcare, and justice will exacerbate social inequality and further cause other social problems.

## 3  Implications of the XAI for governance

XAI is not a new issue. XAI has been studied regarding the early intelligent systems since the early 1970s, and AI explainability has regained much attention in more recent years due to the pervasiveness of AI and the increasing complexity of AI algorithms in the era of neural networks. It is no longer a purely technical issue but is closely related to business interests, ethics, and legislative regulation. Explainability of AI is not the ultimate goal, but a means and a prerequisite for achieving other objectives such as algorithmic fairness and accountability. For example, "one can only make legal arrangements for what is already understood", and explainability is the basis for judging the legal liability of AI. In its 2022 report on the development of XAI, Tencent Research Institute concludes that the significance of interpretability of AI is to help users increase confidence and trust, prevent bias, promote algorithmic fairness, demonstrate that AI systems meet regulatory standards or policy requirements, drive improvements in system design, and help with risk assessment. This is in line with the Royal Society in its policy brief that AI explainability can help promote trust, prevent bias, comply with regulations and policies, and promote human engagement.

For the time being, there are two technical perspectives on XAI: an understanding of the algorithm itself, i.e., a global explanation of the function of the model system, and an understanding of the results of the algorithm, i.e., a local explanation of the specific output of the algorithm. Currently, there is no unified view in the academic and law-making communities on the circumstances in which these two perspectives should come into play. Local explanations help to understand, question, or correct individual cases, but lack the ability to verify the consistency of the AI system. On the other hand, global explanation helps to verify consistency, but it risks revealing trade secrets, thus hindering innovation. In addition, global explanations are technically more difficult to obtain, especially for complex models, such as deep neural networks. Therefore, XAI research needs to consider different user groups and address different explanations needs in order to truly achieve understandable, trustworthy, and controllable AI in practice.

## 4  Global Advances In The Governance Of XAI

Amidst the emerging consensus on AI governance, various sectors are exploring the benefits that explainability can bring. Governments, international intergovernmental organizations, industry standard-setting bodies, and multinational technology companies are all actively exploring the different possibilities of explainability for algorithmic governance.

### 4.1  The United States

As an AI superpower, the U.S. has taken a stance of emphasizing innovation and prudent governance of AI, and uniform laws and regulations have not been established at the federal level. Such an emphasis on technological innovation and decentralized governance also extends to the field of XAI. The Defense Advanced Research Projects Agency launched the XAI program in 2017 with the goal to enable end users to better understand, trust, and effectively manage AI systems. The scholars in this project believed that the measurement of explanation effectiveness is a major challenge of fundamental research on XAI and future research should learn more from sociology, psychology, and other disciplines. In 2021, the National Institute of Standards and Technology proposed four principles for explainable artificial intelligence (XAI): systems deliver accompanying evidence or reason(s) for all outputs; systems provide explanations that are understandable to individual users; the explanation correctly reflects the system's process for generating the output; the system only operates under conditions for which it was designed or when the system reaches sufficient confidence in its output.

## 4.2 The European Union

In comparison to the U.S., where official efforts to explore XAI have mostly come from non-regulatory agencies, the EU has adopted its consistent vigorous regulatory stance on XAI. To date, the regulation of algorithmic explainability in the EU has gone through a process of moving from policy proposals to enforced rules. In 2018, the European Commission published a communication on "Artificial Intelligence in Europe" calling for the EU to lead the way in developing AI on a fundamental rights framework. The same year in June, the European Commission appointed a High-Level Expert Group (AI HLEG)to advise on its AI strategy and draft ethical guidelines. On 8 April 2019, the AI HLEG presented Ethics Guidelines for Trustworthy Artificial Intelligence. The guidelines listed both transparency and explainability of algorithms as one of the 7 key requirements for achieving trustworthy AI, emphasizing the critical role of XAI in building human trust in algorithms. In 2020, the AI HLEG presented their final Assessment List for Trustworthy Artificial Intelligence, which translated 7 principles into an accessible and dynamic checklist to guide AI developers and deployers in implementing these principles in practice.

In 2019, European Parliament's Science and Technology Options Assessment (STOA) Panel published "A governance framework for algorithmic accountability and transparency", in which AI transparency and explainability are conceived as instruments to achieve AI fairness and governance and called upon the EU to establish regulatory mechanisms and legislative frameworks as soon as possible. While neither the guidelines nor the framework are legally mandatory, they both reflected the EU's top-down approach, which emphasizes governance.

In April of 2021, the European Commission submitted its proposal for a European Union regulatory framework on AI. The AI Act represents the first attempt globally to make specialized legislation for AI regulation. The AI Act mandates that high-risk AI systems be 'sufficiently transparent to enable users to interpret the system's output and use it appropriately'. The transparency requirements for AI systems in the Artificial Intelligence Act are intended to facilitate user understanding and are closely linked to the technical view of algorithmic explainability. The explanation requirements can also be found in the EU Platform-to-Business Regulation. This regulation requires online platforms and search engines to provide an easily and publicly available explanation of their ranking algorithms.

## 4.3 China

Although there is not yet legislation specifically for AI, China has now initially established a soft governance framework including algorithm explainability specifications. The National New Generation Artificial Intelligence Governance Specialist Committee released Ethical Norms for New Generation Artificial Intelligence. In Article 12, it is stated that 'in the algorithm design, implementation, and application stages, improve transparency, explainability, comprehensibility, reliability, and controllability'. In short, AI should be explainable. The Cyberspace Administration of China released the draft "Internet Information Service Algorithmic Recommendation Management Provisions". It requires algorithmic recommendation service providers to optimize the transparency and understandability of search, ranking, selection, push notification, and to publicize the basic principles, purposes, motives, and operational mechanisms in a suitable manner. Overly rigid requirements for explainability may hinder evolving AI technologies and place burdens and uncertainties on related business practices. China's current governance of AI is still centered on technological progress, with the bottom line of ensuring safety and controllability, which requires a balance between innovation and the ethics of AI as represented by explainability.

## 4.4 Tech Giants

As the explainability of algorithms has come to the attention of the public, developers and regulators, tech giants such as Google, Microsoft and IBM have begun to invest substantially in XAI research. Model cards released by Google represent its most recent advances in XAI. In the form of visual documents, model cards provide users with a detailed overview of a model's basic operation mechanism, suggested uses and limitations. The target users of model cards are developers specialized in AI. Another tool from Google, 'What-if', offers the possibility of understanding machine learning models to the users without programming skills. The 'What-if' tool can be accessed directly as a web page, where users can explore the results of algorithm on their own through an interactive interface. IBM proposed the AI Explainability 360 toolkit and AI FactSheets 360 to support the explainability of AI models. IBM's research matches different XAI methods to the specific explainability goals faced by different industries and users, providing them with unique explainable reports based on their specific needs, thus encouraging the wider use of XAI techniques at all levels and in different sectors. Although, some research progress has been made in XAI recently, the XAI tools mentioned above are still in the testing phase. Since XAI research is at the frontier of AI research, it requires significant corporate resources and is difficult to transform directly into economic benefits. Tech companies also conduct XAI research mostly out of pressure on public opinion and future regulation. Without timely and mandatory regulation, it remains unclear how long voluntary corporate investment in XAI research will last and to what extent it will translate into products that benefit users.

### 4.5 International cooperation in promoting interpretable XAI

The development and implementation of AI is beyond country borders, and it is of great importance to deliberate on AI innovation and governance at a global collaborative level. In the past few years, international dialogue and cooperation on governance concerning the explainability and ethics of AI has gradually increased, and is mainly led by international intergovernmental organizations. In May 2019, the member states of the Organization for Economic Cooperation and Development (OECD) approved the OECD AI Principles, with transparency and explainability as one of the five basic principles. The OECD encourages its member states' AI actors to provide meaningful information, appropriate to the context, and consistent with the state of art: to foster a general understanding of AI systems; to make stakeholders aware of their interactions with AI systems, including in the workplace; to enable those affected by an AI system to understand the outcome, and to enable those adversely affected by an AI system to challenge its outcome based on plain and easy-to-understand information on the factors, and the logic that served as the basis for the prediction, recommendation or decision. In November 2021, the Recommendation on the Ethics of Artificial Intelligence, the first ever global standard on the ethics of AI, was adopted by 193 member states of UNESCO. The Recommendation defines transparency and explainability as one of the top 10 principles for AI. This symbolizes that the development of XAI has become a broad consensus in the world at the governmental level. However, the international collaboration on algorithmic explainability is still in the negotiation stage, and the extent to which the consensus on developing algorithmic explainability will become a unified international standard is still dependent on the future global collaboration.

## 5  Discussion

The explainability of AI has transformed from a purely technical issue to a complex issue closely related to algorithmic governance and algorithmic security. The lack of explainable AI (XAI) brings adverse effects that can cross all economic classes and national borders. Despite efforts in governance, technical, and policy exchange have been made in XAI by multiple stakeholders, including the public sector, enterprises, and international organizations, respectively. XAI is still in its infancy. Future applications and corresponding regulatory instruments are still dependent on the collaborative engagement of all parties.

Work in progress A PREPRINT